# Ultrafast Photoinduced Formation of Metallic State in a Perovskite-type Manganite with Short Range Charge and Orbital Order

Yoichi Okimoto[1,*], Hiroyuki Matsuzaki[2], Yasuhide Tomioka[1], Istvan Kezsmarki[3,**], Takeshi Ogasawara[1], Masakazu Matsubara[1], Hiroshi Okamoto[1,2], and Yoshinori Tokura[1,3]

[1]*Correlated Electron Research Center (CERC), National Institute of Advanced Industrial Science and Technology (AIST), Tsukuba, 305-8562 Japan*
[2]*Department of Advanced materials Science, University of Tokyo, Kashiwa, 277-8561, Japan*
[3]*Department of Applied Physics, University of Tokyo, Tokyo 113-8656, Japan*

Femtosecond reflection spectroscopy was performed on a perovskite-type manganite, $Gd_{0.55}Sr_{0.45}MnO_3$, with the short-range charge and orbital order (CO/OO). Immediately after the photoirradiation, a large increase of the reflectivity was detected in the mid-infrared region. The optical conductivity spectrum under photoirradiation obtained from the Kramers-Kronig analyses of the reflectivity changes demonstrates a formation of a metallic state. This suggests that ferromagnetic spin arrangements occur within the time resolution (~200 fs) through the double exchange interaction, resulting in an ultrafast CO/OO to FM switching.

Recently, researches on the transient changes of electronic phases in solids have drawn considerable attention, as accelerated by the recent developments of femtosecond (fs) laser technology. The irradiation of an ultrashort laser pulse often produces a new phase hidden in a material that hardly appears by controlling external parameters such as temperature, pressure and magnetic fields in a steady state. This phenomenon is called photoinduced phase transition (PIPT)[1], and several interesting examples using various optical techniques have been reported, in particular for the strongly correlated electron systems (SCESs)[2-3].

Perovskite-type manganite is a typical example of SCESs[4] and of current interest due to the significant electronic phase transitions, as exemplified by the melting of the real space charge order (CO) and orbital order (OO) into the ferromagnetic metallic (FM) state by the external stimuli such as magnetic/electric field, x-ray irradiation, impurity doping, and so on [5]. Among them, photoirradiation can also drive the phase transition, which was first demonstrated in a CO/OO manganite, $Pr_{0.7}Ca_{0.3}MnO_3$[6]. This is a quasi-static phase transition from the CO/OO state to the metallic state, induced by the irradiation of a ns laser pulse in cooperation with current injection. This can be viewed as an interesting example of PIPT and arouse many researches on the PIPTs in the maganites using ns laser light[7]. However, the mechanism of the PIPT has remained unclear due to the lack of the information about the ultrafast dynamics in a fs time scale. Some pioneering works using the fs laser spectroscopy reported the ultrafast melting of the CO/OO in maganites[8-10], but a PIPT to the FM state has not yet been confirmed. Therefore, the time scale of creation and annihilation of the FM domains have remained to be clarified.

In this Letter, we performed a time-resolved reflection spectroscopy on a CO/OO manganite, $Gd_{0.55}Sr_{0.45}MnO_3$, using a fs pump-probe (PP) method. The electronic phase diagram associated with $R_{0.55}Sr_{0.45}MnO_3$ as a function of the averaged ionic radius $<r>$ of $R_{0.55}Sr_{0.45}$ is shown in Fig. 1(a) [11]. The decrease of $<r>$ suppresses the one-electron bandwidth of $e_g$ electrons and hence decreases Curie temperature ($T_c$) in the FM state as seen for $<r>$=1.4 Å to 1.32 Å. With further decreasing $<r>$, the FM state disappears and the CO/OO state instead appears as shown in Fig. 1(a). A recent x-ray diffraction study indicates that the correlation length ($\xi$) of the CO/OO states in this region of the phase diagram labeled SGI in Fig. 1(a) is quite short (10-20Å)[11], and the magnetic measurement suggests a spin glass transition at $T_{SG}$~50K[11,12]. This is because the difference of the ionic size between $Sr^{2+}$ and the trivalent rare earth ions is so large that the resultant local randomness suppresses an evolution of long range CO/OO[12]. As seen in the phase diagram, the ground state of $Gd_{0.55}Sr_{0.45}MnO_3$ is the short range CO/OO state located on the verge of the phase boundary. We also show in Fig. 1(b) the temperature dependence of resistivity ($\rho$) in magnetic fields of 0 T and 7 T for $Gd_{0.55}Sr_{0.45}MnO_3$. The $\rho$-$T$ curve at 0 T indicates an insulating behavior (d $\rho$/d$T$<0). In a magnetic field of 7 T, by contrast, the $\rho$-value significantly decreases and the FM state is realized. This also implies the minimal energy difference between the short-range CO/OO state and the FM state. This fact as well as the small value of $\xi$ in the CO/OO state suggests that $Gd_{0.55}Sr_{0.45}MnO_3$ is a good candidate for the study of photoinduced transition from the CO/OO to the FM state. Hereafter, we report on the ultrafast dynamics in the CO/OO to FM transition in $Gd_{0.55}Sr_{0.45}MnO_3$.

A single crystal of $Gd_{0.55}Sr_{0.45}MnO_3$ was grown by the floating zone method[11]. The reflectivity spectrum ($R(\omega)$) was measured using a Fourier transform-type interferometer (0.01-0.7 eV) and grating monochromator (0.6-5 eV). The reflectivity variation by irradiating a fs laser pulse ($\Delta R/R$) was detected by a PP method using a Ti:sapphire regenerative amplifier system (pulse width of ~100 fs, repetition rate of 1 kHz, and photon energy of 1.55 eV) as a light source. The amplified light was split into two light paths. One was used as a pump light for the photoexcitation. The other whose frequency was converted from 0.12 to

2.1 eV by an optical parametric amplifier (OPA) was used as a probe light to investigate $R(\omega)$ after the photoexcitation.

In Fig. 1(c), we show the $R(\omega)$ spectrum in $Gd_{0.55}Sr_{0.45}MnO_3$ at 10 K, which is characteristic of the CO/OO insulating state of the manganites. The broad peak at 0.2-2 eV is due to the transition from $Mn^{3+}$ to $Mn^{4+}$ in the $e_g$ orbital sector[13]. The spiky structures below 0.08 eV are attributable to the optical phonon modes. In Fig. 1(d), we show the $\Delta R/R$ spectra at several typical delay times ($t_d$) of the probe pulse relative to the pump pulse. The photon energy of a pump pulse (1.55 eV) corresponds to the $Mn^{3+}$ to $Mn^{4+}$ transition. The excitation power density of the pump pulse is ~9 mJ/cm$^2$, from which the excitation photon density is evaluated to be ~0.1 photon/Mn site. Immediately after the photoirradiation, *i.e*, at $t_d$~0 ps (filled circles)[14], $\Delta R/R$ increases below 0.7 eV, but decreases above 0.7 eV. The monotonous increase of $\Delta R/R$ with decreasing photon energy and the large increase of the reflectivity ($\Delta R/R$~0.8 at 0.12 eV) signal that the photoirradiation produces a metallic state. Importantly, the initial change of $\Delta R/R$ largely diminishes at $t_d$=1 ps, and the $\Delta R/R$ spectra hardly changes from $t_d$=1 ps (open ciecles) at least to 20 ps (open triangles) throughout the energy region. The spectral shape of $\Delta R/R$ for $t_d$=1 ps suggests that the initially photoproduced metallic state returns to the insulating state within ~1 ps.

To see the transient optical response more quantitatively, we show in Fig. 2 the time evolution of $\Delta R/R$ at selected photon energies of (a) 0.12, (b) 0.6, (c) 0.9 and (d) 1.3 eV with open circles. As expected from Fig. 1(d), the time profiles are composed of two components; the fast component decaying within $t_d$ ~1 ps and the less time-dependent slow component being dominant for $t_d$ > 1 ps. It is reasonable to consider that the former resulted from the photoinduced metallic states and their decay, and the latter from the increase of the lattice temperature induced by the decay of the initial photoinduced metallic states. On this basis, the time profiles can be fitted by the following function, $f(t) = I_1 \exp(-t/\tau_1) + I_2 (1-\exp(-t/\tau_1))$. The first term denotes an exponential decay and the second a constant component that increases with the relaxation of the first term. The actual fitting was performed with the convolution between f(t) and the response function of the present fs PP system. The temporal width of the laser pulse is slightly changed with the probe wavelength. Therefore, the magnitude of the time resolution in the response function is modified depending on the probe wavelength so that the good fitting curves can be obtained. In Fig. 2, we plotted the results of the fitting with black lines, which reproduce well the experimental data. The broken and dotted broken lines show the first and second components in f(t), respectively. According to the analyses, the decay time $\tau_1$ was evaluated to be ~280 fs, indicating the ultrafast decay of the initial photoinduced state. In Fig. 2(e), we show the spectra of the fast component (filled circles) and the slow component (open circles) around 0 ps ($t_d=t_{peak}$), obtained by the fitting procedure. For comparison, we also show the variation of the $R(\omega)$ between 10 and 300 K, *i.e.*, $(R(T=300 K)-R(T=10 K))/R(T=10 K)$ with a solid line. The absolute value as well as the spectral shape of the calculated differential reflectivity is largely different from that in the

$\Delta R/R$ for the fast component (filled circles), whereas roughly similar to the $\Delta R/R$ spectrum for the slow component (open circles). This supports our interpretation that the fast component is attributable to the photoinduced change to the metallic state and the slow component to the decaying process to a charge and orbital disordered state. The rise of the slow component is very fast, determined by the decay time of the metallic state as shown by the fitting procedures. Therefore, it is natural to consider that the electron and lattice system is not completely thermalized for $t_d$= 1ps to 20 ps.

In Fig. 3, we plotted the pump-power dependence of the fast and slow components of $\Delta R/R$ at 0.12 eV by the flled and open circles, respectively. As the pump power increases, the fast component in $\Delta R/R$ increases linearly up to ~3 mJ/cm$^2$. It tends to saturate above ~3 mJ/cm$^2$ and becomes almost constant at around 9 mJ/cm$^2$. By contrast, the slow component is almost proportional to the pump power up to ~9 mJ/cm$^2$. Such a difference in the pump-power dependence also reflects the difference of the photoinduced and the thermal effects.

To clarify the nature of the photoinduced phase more in detail, we evaluate a complex dielectric function in the photoinduced metallic state ($\varepsilon^{PI}$) in terms of Kramers-Kronig (KK) analysis. We can numerically calculate the reflection phase change ($\theta$) of the probe light around 0 ps ($t_d=t_{peak}$) from the reflectivity spectrum (see Fig. 2(e)) considering the KK relation. If the electronic state is uniform in the crystal, a dielectric function can be algebraically obtained from R and $\theta$. In the case of PIPT, however, the number of photons absorbed in the sample exponentially decreases with increase of the distance ($z$) from the surface depending on the penetration depth $d$ of the pump light. In the present case $d$~530 Å[15]. The yields of PIPT will also decrease along $z$ (see Fig. 4(a)). Thus, the reflectivity of the probe light is determined by the dielectric constant ($\varepsilon^{crystal}(z)$) continuously changing with $z$, which can be expressed by the following relation, $\varepsilon^{crystal}(z) = \gamma \exp(-z/d) \varepsilon^{PI} + (1-\gamma \exp(-z/d)) \varepsilon^{CO}$. $\varepsilon^{CO}$ is the original dielectric function of the short-range CO/OO state and $\gamma$ is the efficiency of the photoinduced phase transition (0<$\gamma$<1). Here, we analyze the data for the pump-power density of ~9 mJ/cm$^2$, at which the photoinduced change is saturated (see Fig. 3). Hence we assume that $\gamma$=1. With these conditions, complex refractive index ($r = R^{1/2}\exp(i\theta)$) at $z$=0 obtained by the K-K analysis can be described as a function of $\varepsilon^{PI}$ (i.e., $r = r(\varepsilon^{PI})$). By numerically solving this equation, we can extract $\varepsilon^{PI}$ as well as $\varepsilon^{crystal}(z)$ from the obtained data. We plotted in Fig. 4(b) the value of Im $\varepsilon^{crystal}(z)$ in a plane of the photon energy and $z$. As can be seen, $\varepsilon^{crystal}(z)$ largely changes along $z$ especially in the infrared region.

In Fig. 4(c), we show the numerically calculated optical conductivity spectrum in the photoinduced state ($\sigma^{PI}(\omega) = \omega \text{Im}[\varepsilon^{PI}]/4\pi$) around 0 ps ($t_d=t_{peak}$) with solid circles. For comparison, we plotted linear optical conductivity ($\sigma(\omega)$) of Gd$_{0.55}$Sr$_{0.45}$MnO$_3$ at 10 K with a solid line, which was derived from the R($\omega$) spectrum by the KK analysis (Fig. 1(c)). The broad peak around 1.2 eV is due to the electron transition from the Mn$^{3+}$ to Mn$^{4+}$ site as mentioned above[14]. Just after the photoirradiation ($t_d=t_{peak}$), the spectral weight around the broad peak is decreased and transferred to the lower energy region below 0.6 eV. Such a spectral-weight transfer in a wide energy region is very similar to the temperature variations of $\sigma(\omega)$ in the manganites which undergo the CO/OO insulator to the FM phase transition with the decrease in temperature[13,16,17]. This strongly suggests that the photoinduced state in $Gd_{0.55}Sr_{0.45}MnO_3$ is the FM state. In Fig. 4, we plotted the value of the dc conductivity ($\sigma_{dc}$) at the lowest temperature in 7 T (see Fig. 1(a)), with the solid square on the vertical axis for comparison. The value of $\sigma^{PI}(\omega)$ extrapolated to $\omega=0$ ($\sigma(0)$) is in accord with $\sigma_{dc}$. This also supports the scenario of the photoinduced transition to the FM state.

It is worth noting that the $\sigma^{PI}(\omega)$ spectrum has a broad peak structure at ~0.4 eV and can hardly be regarded as a simple metal expressed in terms of a Drude model. A similar unusual $\sigma(\omega)$ spectrum is observed in manganites such as $Sm_{0.6}Sr_{0.4}MnO_3$ and bilayered manganites, which are ferromagnetic metals with a relatively large residual resistivity[16,17]. The electronic origin of the peak in $\sigma(\omega)$ has not been fully unraveled yet. For the interpretation of the peak, it will be important to consider the intraband transition of the $e_g$ bands composed of $d_{x2-y2}$ and $d_{3z2-r2}$ orbitals[18] and/or the strong orbital correlation in the FM state with the narrow $e_g$-bandwidth[19]. The large residual resistivity in the FM state of $Gd_{0.55}Sr_{0.45}MnO_3$ signals the suppression of the Drude (coherent) component, which would enable us to observe such an intraband absorption.

The $\sigma(\omega)$ spectrum upon photoexcitation in $Gd_{0.55}Sr_{0.45}MnO_3$ indicates that the formation of the FM state occurs very fast, at least within the time resolution in our fs PP system (~200 fs). To the best of our knowledge, a photoinduced insulator-metal transition within a sub-ps time scale has never been demonstrated in the manganites. For the formation of the metallic states, ferromagnetic spin arrangements are indispensable, which should be induced through the double exchange (DE) interaction. The timescale of the formation of the FM state by the DE mechanism can be described as ~$xt$-$J_s$[20], where $x$, $t$ and $J_s$ are the hole concentration, the transfer energy of $e_g$ electrons and the superexchange energy between $t_{2g}$ spins, respectively. If we assume that $t$~0.5 eV[21] and $J_s$~8 meV[20,22], $xt$-$J_s$~ 0.22 eV (~20 fs). This value (~20 fs) is much smaller than the resolution of our system, consistent with the observed ultrafast FM transition.

Finally, we briefly discuss the ultrafast recovery of the photoinduced metallic state (~280 fs), which is another important feature of the present PIPT. In the manganites, the photoinduced melting of the CO/OO states were previously reported in $(Pr,Ca)MnO_3$[8] and $(Nd,Ca)MnO_3$[9] which are the typical CO/OO systems with long range orders, although the formation of the photoinduced metallic states has not been demonstrated. In these systems, the decay time of the photoinduced states seems to be longer than that in the short range CO/OO system, $Gd_{0.55}Sr_{0.45}MnO_3$. This implies that the smaller $\xi$ is important for the ultrafast recovery to the CO/OO state.

In summary, we performed femtosecond pump-probe reflection spectroscopy on the short-range CO/OO manganite, $Gd_{0.55}Sr_{0.45}MnO_3$, and observed the large and ultrafast changes of the reflectivity spectrum. The detailed analyses of the transient reflectivity spectra by using the Kramers-Kronig relation revealed that the optical conductivity spectrum for the photoinduced state is very similar to that observed in the ferromagnetic metallic states of the manganites, suggesting that the ultrafast optical switching from the CO/OO state to the FM state. The formation of the FM state and its ultrafast recovery to the CO/OO state is considered to be characteristic of the short-range nature of CO/OO in this manganite.

The authors thank S. Ishihara, N. Nagaosa, and S. Iwai for discussions. This work was supported in part by Grant-In-Aids for Scientific Research from the MEXT, Japan.

*Present address: Department of Materials Science, Tokyo Institute of Technology, Meguro-ku, Tokyo 152-8551, Japan.
**Present address: Department of Physics, Budapest University of Technology and Economics, 1111 Budapest, Hungary.

**Figure captions**

**Fig.1 (color online)**

(a): The electronic phase diagram in $R_{0.55}Sr_{0.45}MnO_3$. (b): Temperature dependence of resistivity in $Gd_{0.55}Sr_{0.45}MnO_3$ in a magnetic field of 0 T and 7 T. (c): Reflectivity spectrum in $Gd_{0.55}Sr_{0.45}MnO_3$ at 10 K. (d): Photoinduced reflectivity change ($\Delta R/R$) spectra at selected delay times ($t_d$). The pump energy is denoted by an arrow.

**Fig.2 (color online)**

Time evolution of $\Delta R/R$ around 0.12 eV (a), 0.4 eV (b), 0.9 eV (c), and 1.3 eV (d). Black lines are the result of the fitting. Broken and dotted broken lines denote two components in f(t), fast and flat component, respectively (see text).

(e): Spectra of the fast component (filled circles) and the flat component (open circles) around 0 ps in the total value of $\Delta R/R$. The bold line shows a difference spectrum of the reflectivity between 10 K and 300 K.

**Fig.3**

Excitation energy dependence of the fast component (filled circles) and the flat component (open circles) at $t_d\sim t_{peak}$ in $\Delta R/R$ around 0.12 eV.

**Fig.4 (color online)**

(a): A schematics that accounts for the spatial variation of the dielectric function, $\varepsilon^{crystal}(z)$ (=

exp($-z/d$) $\varepsilon^{PI}$ + (1-exp($-z/d$)) $\varepsilon^{CO}$). (b): A contour plot of Im[$\varepsilon^{crystal}(z)$] in the plane of $z$-$h\omega$. (c): Numerically calculated optical conductivity ($\sigma(\omega)$) in the photoinduced state at $t_d=t_{peak}$ in Gd$_{0.55}$Sr$_{0.45}$MnO$_3$. A solid curve is a linear $\sigma(\omega)$ in the same crystal at 10 K.

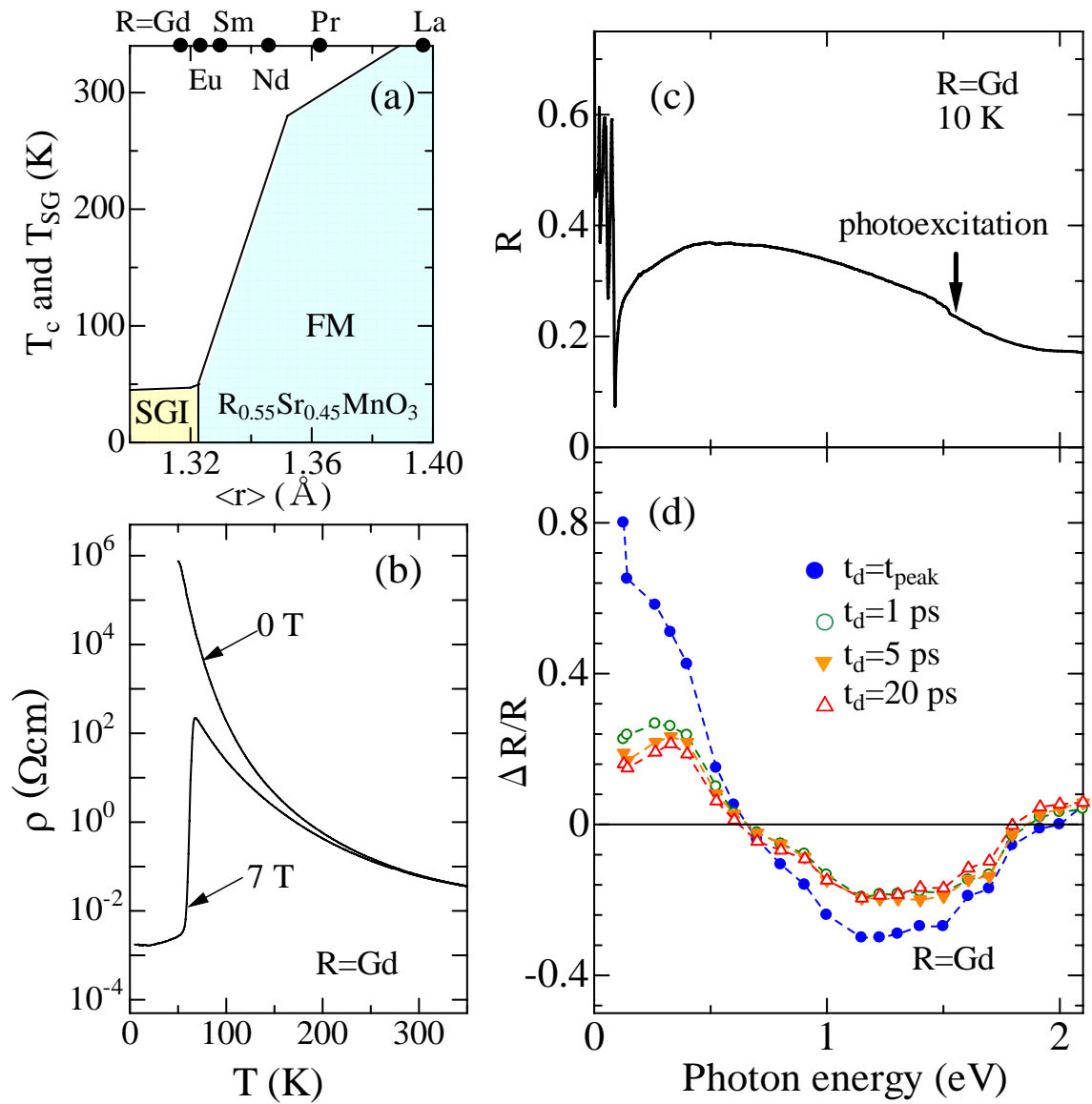

Fig.1 Okimoto *et al.* (color online)

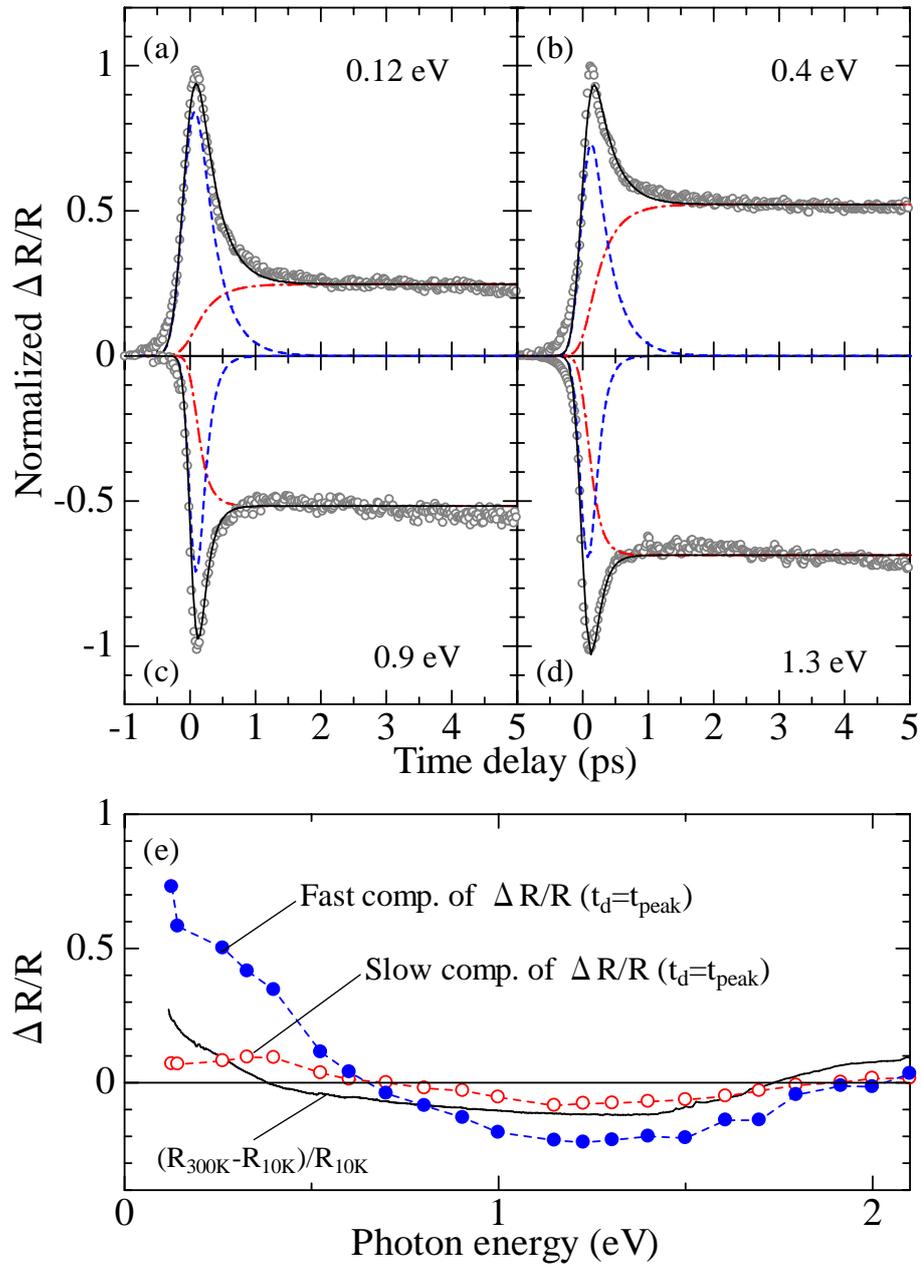

Fig.2 Okimoto *et al.* (color online)

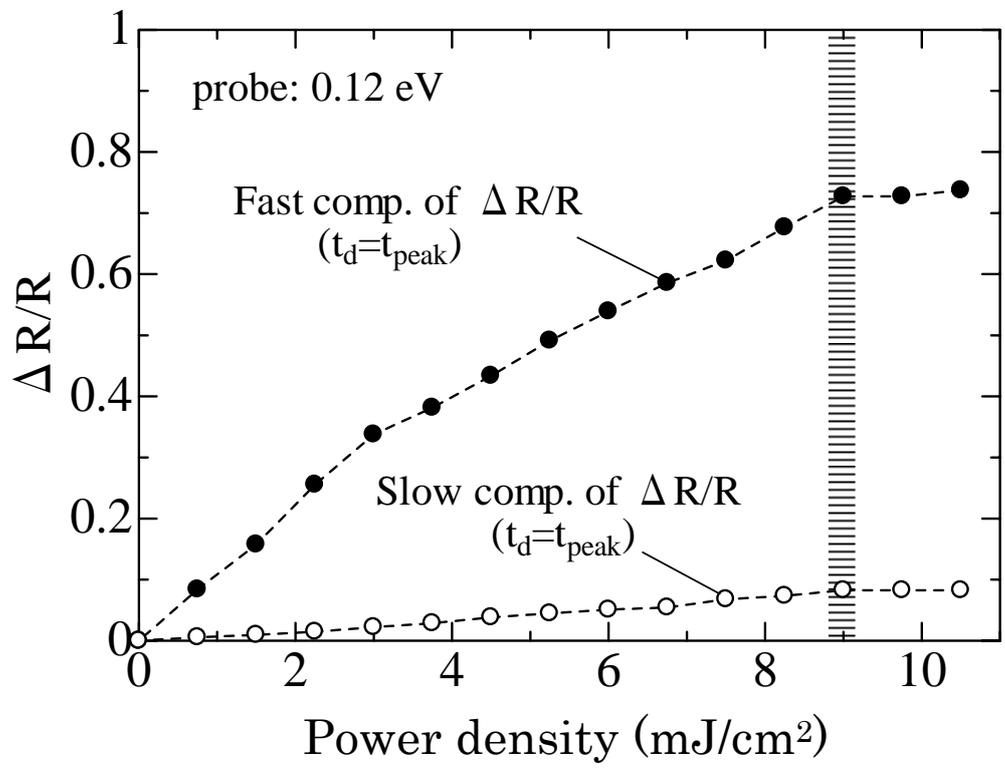

Fig.3 Okimoto *et al.*

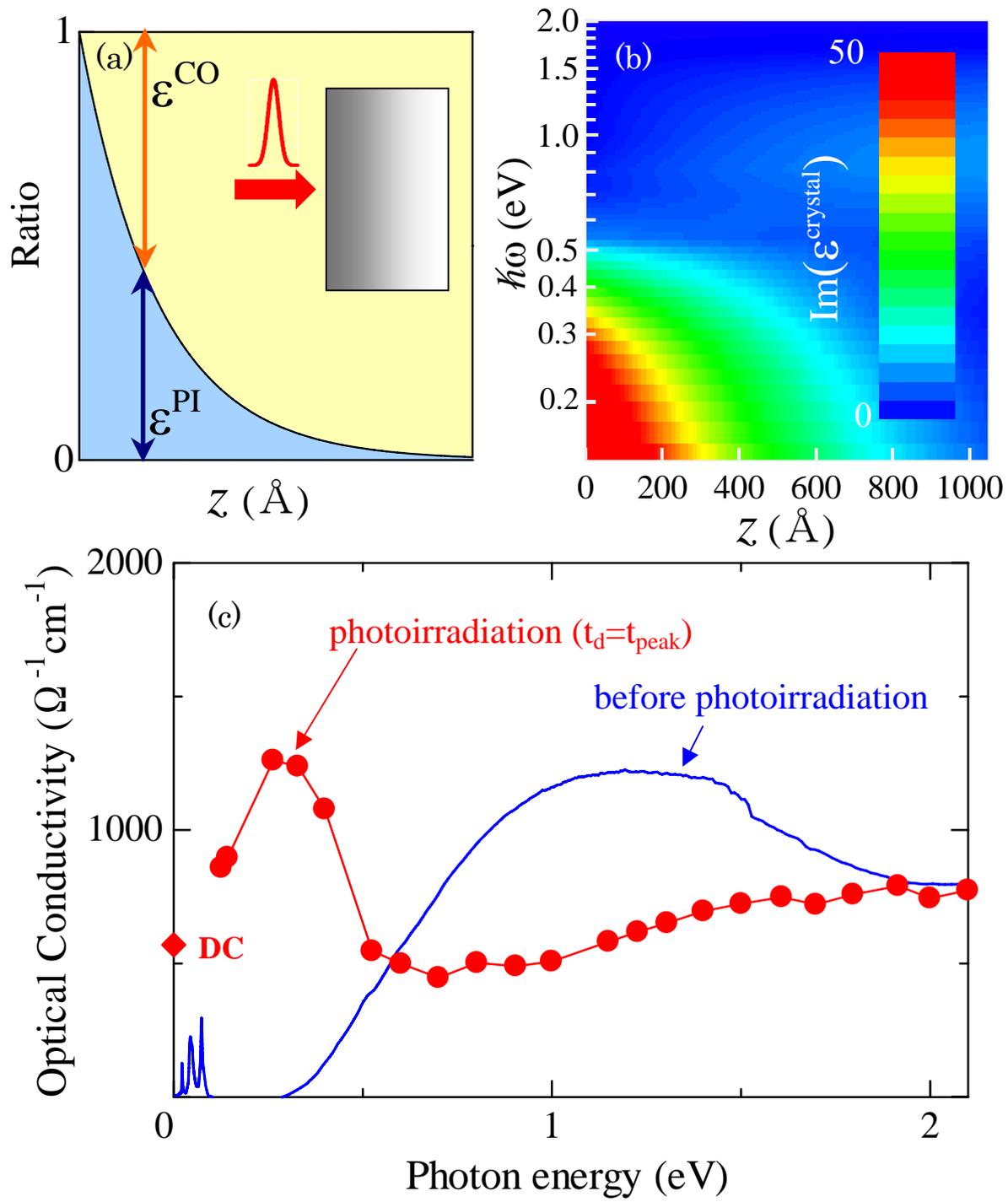

Fig.4 Okimoto *et al.* (color online)